%% file: emufog.tex
\documentclass[conference]{IEEEtran}
\IEEEoverridecommandlockouts

\usepackage{color}

\ifCLASSINFOpdf
  \usepackage[pdftex]{graphicx}
  \graphicspath{/graphics}
   \DeclareGraphicsExtensions{.pdf}
\else
\fi
\usepackage{overpic}

\usepackage{url}
\usepackage{xspace}
\newcommand{\sys}{EmuFog\xspace}
\usepackage{tikz}

\usepackage{algorithm}
\usepackage{algpseudocode}
\begin{document}
%
\title{\sys : Extensible and Scalable Emulation of Large-Scale Fog Computing Infrastructures\thanks{This work was funded in part by DFG grant RO 1086/19-1 (PRECEPT), an NSF CPS program Award \#1446801, GTRI’s IRAD program, and a gift from Microsoft Corp.}}

\author{\IEEEauthorblockN{Ruben Mayer, Leon Graser}
\IEEEauthorblockA{Institute of Parallel and Distributed Systems\\
University of Stuttgart, Germany\\
Email: ruben.mayer@ipvs.uni-stuttgart.de, \\leon.graser@gmail.com }  
\and
\IEEEauthorblockN{Harshit Gupta, Enrique Saurez, \\Umakishore Ramachandran}
\IEEEauthorblockA{Georgia Institute of Technology\\
Atlanta, Georgia, USA\\
Email: \{harshitg, esaurez, rama\}@gatech.edu}}


%


\maketitle
\begin{tikzpicture}
\begin{scope}[overlay]
\footnotesize
\node[text width=40cm] at ([yshift=-18.7cm,xshift=9cm]current page.south) {\copyright 2017 IEEE. Personal use of this material is permitted. Permission from IEEE must be obtained for all other uses,  in any current or future media, \newline including reprinting/republishing this material for advertising or promotional purposes,  creating new collective works, for resale or redistribution \newline to servers or lists, or reuse of any copyrighted component of this work in other works. \newline
This is the authors' version of the work. The definite version is published in Proceedings of 2017 IEEE Fog World Congress (FWC '17).};
\end{scope}
\end{tikzpicture}

\begin{abstract}

The diversity of Fog Computing deployment models and the lack of publicly available Fog infrastructure makes the design of an efficient application or resource management policy a challenging task. Such research often requires a test framework that facilitates the experimental evaluation of an application or protocol design in a repeatable and controllable manner. In this paper, we present EmuFog---an extensible emulation framework tailored for Fog Computing scenarios---that enables the \textit{from-scratch} design of Fog Computing infrastructures and the emulation of real applications and workloads. EmuFog enables researchers to design the network topology according to the use-case, embed Fog Computing nodes in the topology and run Docker-based applications on those nodes connected by an emulated network. Each of the sub-modules of EmuFog are easily extensible, although EmuFog provides a default implementation for each of them. The scalability and efficacy of \sys are evaluated both on synthetic and real-world network topologies.
\end{abstract}


%
\IEEEpeerreviewmaketitle

\input{content/introduction.tex}
\input{content/related.tex}

\input{content/design.tex}

\input{content/evaluation.tex}

\input{content/conclusion.tex}



%
\IEEEtriggeratref{9}
\bibliographystyle{plain}
\bibliography{references}

\end{document}

%% file: content/introduction.tex
\section{Introduction} 
\par The Fog Computing paradigm has emerged to address the issues of high latency and low throughput connections to applications running in remote data centers by bringing compute, storage and networking services close to the users. Fog Computing can be viewed as a non-trivial extension of Cloud Computing, thus creating a continuum of resources extending from the network edge to data centers at the core of the network. 

\par A number of proposals have discussed architectures for implementing Fog Computing, including approaches based on micro-data centers \cite{microdc}, smart gateways \cite{smartgw}, and even multi-level resources \cite{openfog, hong2013mobile, foglets}. Each of these architectures have typical characteristics which have an impact on the optimal design and performance of applications built for them. To design resource management systems, the Fog Computing architecture needs to be taken into account. Furthermore, applications and protocols themselves need to be tested in a repeatable and controllable manner, so that they can be debugged and tuned according to the target Fog infrastructure before actual deployment. Due to the lack of easily accessible public Fog infrastructures, a test environment facilitating the development and deployment of applications becomes a necessity. 

\par This paper aims to fill the void of a test environment for Fog Computing through \sys\footnote{EmuFog is open source: https://github.com/emufog/emufog}, an extensible emulation framework built on top of MaxiNet \cite{6857078} (MaxiNet is a multi-node extension of the popular network emulator Mininet \cite{Lantz:2010:NLR:1868447.1868466}). Compared to simulation and real-world testbeds, emulation supports both repeatable and controllable experiments with real applications. 
\par \sys allows developers to design the Fog infrastructure topology in an extensible manner through a two-step process. The first step builds a network topology of switches/routers, which in \sys can be done by employing network topology generators such as BRITE \cite{Medina:2001:BAU:882459.882563} or by importing real-world topology datasets, e.g., from CAIDA \cite{caida}. The next step is to place Fog Computing nodes in the generated network topology, which the developer can customize according to different placement policies and by the specification of Fog node capabilities and expected workload. Once the complete Fog Computing topology is generated, it is fed into MaxiNet for emulation. Although \sys provides a standard implementation for fog computing topology generation, the developer can fully customize it to suit her needs.

\par The outline of the paper is as follows. Section \ref{sec:background} discusses the background of test frameworks for Fog Computing. Section \ref{sec:design} presents the design of the \sys framework and provides implementation details. In Section \ref{sec:evaluation}, the performance and efficacy of \sys are evaluated.

%% file: content/related.tex
\section{Background}
\label{sec:background}
\subsection{Fog Computing}
Fog Computing is a non-trivial extension of the popular Cloud Computing paradigm that brings compute, storage, control and networking services close to the users. Here, the phrase ``\textit{close to users}" refers to any point between the data centers at the core and the users at the edge of the network \cite{openfog}. Fog infrastructure spans across several network domains -- all working in coordination as an ecosystem for enhanced application delivery.

 
\subsection{Test Environments for Fog Computing} The highly heterogeneous and geo-distributed nature of Fog Computing, coupled with the lack of a real testbed or commercial service, has been a driving factor for research in test environments for Fog Computing. Gupta et al. proposed \textit{iFogSim} \cite{ifogsim}, a simulation toolkit for evaluating application design and resource management techniques in Fog Computing ecosystems. The toolkit performs Discrete Event Simulation (DES) allowing users to simulate Fog Computing infrastructure and run simulated applications on it to measure the performance in terms of latency, energy consumption and network usage.

\par However, simulations make a number of simplifications that may not always hold true, especially with an infrastructure as dynamic as Fog Computing, due to which we chose to develop an emulation framework. Existing network emulators like MaxiNet~\cite{6857078} and CORE \cite{core} allow users to emulate network devices and hosts, thereby making it possible to perform repeatable and configurable experiments while also testing with real applications. In principle, in a network emulation framework, a user can setup experiments to run applications in a Fog Computing infrastructure; the effort of doing that, however, is quite high as these frameworks are very general purpose. In particular, the placement of Fog nodes in the network would need to be performed by hand, which is infeasible for large scenarios. Hence, there is a need for an emulation framework custom-made for Fog Computing that diminishes the users' effort in experimentation.

In this regard, MaxiNet is a distributed network emulator which has been implemented by extending Mininet~\cite{Lantz:2010:NLR:1868447.1868466}. Contrary to the single-node emulation of Mininet, MaxiNet is able to span an emulated network over several machines, enabling it to emulate networks with several thousand nodes on a few physical machines. The authors of MaxiNet show that MaxiNet is able to emulate a datacenter network with 3200 hosts on just 12 physical machines. MaxiNet's ability to support such large-scale emulations is crucial to the performance of Fog Computing emulations, owing to the large scale and distributed nature of Fog Computing infrastructures. Hence, we use MaxiNet as a basis of \sys, extending it with native support for Fog Computing scenarios.


%% file: content/design.tex
\begin{figure*}
     \centering
     \includegraphics[width=0.82\linewidth]{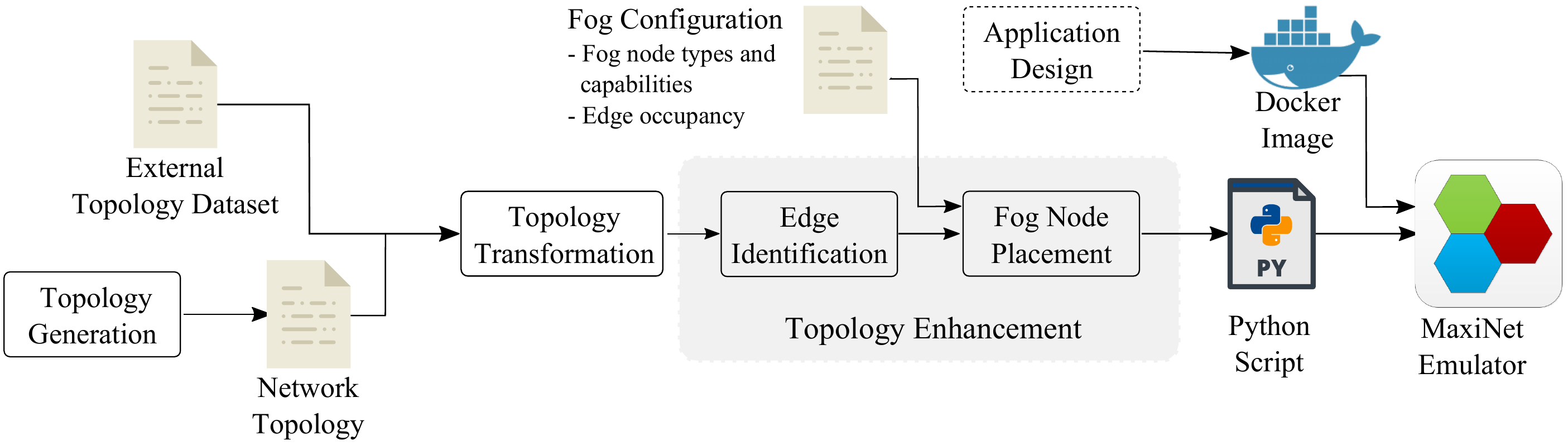}
     \vspace{-0.2cm}
     \caption{Workflow of setting up an \sys emulation. Rectangles with solid outline represent functionalities inside \sys that can be adapted by the user to her needs. \sys already provides standard implementations of those functionalities that are suitable for a large range of Fog Computing scenarios.}
     \label{fig:workflow}
     \vspace{-0.3cm}
 \end{figure*}

\section{Design and Implementation}
\label{sec:design}
In this section, we introduce the design of \sys and provide details of the implementation of the main components. In particular, we detail the implemented algorithms for placing Fog nodes in an imported network topology.

\subsection{Design Objectives}
\sys implements the following design objectives.

\begin{itemize}
\item \textbf{Scalability for large-scale topologies} : \sys can emulate large network topologies allowing the developer to study large-scale Fog Computing scenarios.
\item \textbf{Emulation of real applications and workloads} : \sys makes it easy for developers to package their applications and run them in the emulated environment. This makes it convenient to test the designed Fog Computing policies and applications against real workloads at scale.
\item \textbf{Extensibility} : All components of \sys are extensible and replaceable by custom-built components that suit the scenario to be emulated or the policies to be evaluated.
\end{itemize}

\subsection{Workflow of Fog Computing Emulations}

The workflow when performing Fog Computing emulations in \sys is depicted in Figure \ref{fig:workflow}. It consists of 4 main steps:

\begin{enumerate}
	\item \textbf{Topology Generation:} A network topology is generated by a network topology generator, such as BRITE \cite{Medina:2001:BAU:882459.882563}. The network topology can also be loaded from a file, which allows for including real-world topology datasets.
    \item \textbf{Topology Transformation:} In \sys, the network is represented as an undirected graph of network devices (routers) connected by links with certain latency and throughput. Network devices are grouped into Autonomous Systems (AS). Hence, the generated or imported network topology is translated into the network topology model of \sys.
    \item \textbf{Topology Enhancement:} The network topology is enhanced with Fog nodes. In doing so, two sub-steps are performed: First, the edge of the network topology is determined. Second, Fog nodes are placed in the network topology according to a placement policy. To this end, the user specifies in a \emph{Fog configuration} file the type of Fog nodes and their computational capabilities at a high level. Further, the user specifies how many clients he expects at the network edge connecting to the application deployed in the Fog infrastructure.
    \item \textbf{Deployment and Execution}: The enhanced network topology is deployed in the emulation environment. In particular, Fog nodes are placed in the emulated network, and the application components, provided as Docker containers, are deployed on the Fog nodes. 
\end{enumerate}

This way, application developers can easily evaluate their Fog Computing applications in different Fog environments. For instance, they can test the behavior of the application when Fog nodes are deployed very close to the edge of the network, or in Fog Computing environments where Fog nodes are placed rather far away from the edge. As there are many different views on the future design of the Fog Computing continuum~\cite{openfog}, running application evaluations in flexible Fog Computing environments is of great help to the application developers. 

Furthermore, a Fog Computing designer can evaluate the implications of different Fog node placement policies in the network topology and answer ``what if'' questions. For instance, she can analyze the latency and cost implications for supporting a number of clients for a specific Fog configuration. This can provide useful hints when laying out the Fog infrastructure.

All steps in the workflow of an \sys experiment can be implemented according to the user's needs. However, \sys already provides a set of implementations that are suitable to serve a large set of Fog Computing emulation scenarios. \sys provides a topology generation component to generate Internet scale topologies, which is based on the BRITE network topology generator \cite{Medina:2001:BAU:882459.882563}. Further, an adapter is provided to translate BRITE network topologies as well as real-world topologies from the CAIDA Internet Topology Data Kit (ITDK) to the network topology model of \sys. Those components are rather simple; hence, we omit further details in this paper. In the following, the implementation of the topology enhancement components is discussed in more detail.

\subsection{Topology Enhancement}

The network topology is enhanced with Fog Computing nodes based on particular placement algorithms. Malandrino et al. \cite{placement} discuss the placement of Fog nodes in a network and evaluate placements in terms of server utilization and application latency, showing that the best Fog placement strategy will depend on individual network operator's deployment strategy and geographic workload distribution. To this end, \sys provides an extensible component to perform Fog node placement in the network topology. 

In \sys, we implemented a novel latency-based Fog node placement policy, that aims to keep a latency bound between the clients connecting at the network edge and the closest Fog node. This can be extended by the user to consider other metrics such as bandwidth. 
In the following, we provide details of the edge node identification and Fog node placement algorithms that implement the latency-based placement policy.

\subsubsection{Identification of Edge Routers}\label{sec:edgeId}

\emph{Edge routers} refer to the routers in an AS that serve as access points for clients to connect to the network, while the rest of the routers in an AS will be termed backbone routers in this paper. We assume that a client connects to one fixed edge router to join the network. In order to support Fog node placement policies that take into account the latency between the clients and the Fog nodes, the location of edge routers is a crucial factor.

\begin{algorithm}
\small
	\caption{Backbone Connection Algorithm}
	\label{alg:connectBackbone}
	\begin{algorithmic}[1]
		\Procedure{ConnectBackbone}{$B,G=(V,E)$}
			\State $b \gets b \in B$
			\State $Q \gets \{b\}$
			\While{$Q \neq \{\}$} 
				\State $v \gets Q.\Call{Dequeue}{ }$
				\If{$v \in B \wedge v.parent \in V \setminus B$}
					\State $p \gets v.parent$
					\While{$p \in V \setminus B$}
						\State $B \gets B \cup \{p\}$
						\State $p \gets p.parent$
					\EndWhile
				\EndIf
				\ForAll{$n \in N_1{(v)}$}
					\If{$n \notin Q$}
                    	\State $n.parent \gets v$
						\State $Q.\Call{Enqueue}{n}$
                     \Else
						\If{$v \in B \wedge n.parent \in V \setminus B$}
							\State $n.parent \gets v$
						\EndIf
											
					\EndIf
				\EndFor
			\EndWhile
			\State \textbf{return} $B$
		\EndProcedure
	\end{algorithmic}
\end{algorithm}

\par The algorithm for identifying edge routers starts from a state where all routers in an AS are marked as edge routers, gradually moving routers from the edge router set to the set of backbone routers. The algorithm proceeds in 3 steps, described as follows.
\par \textbf{Step 1}: Routers connecting different ASs are marked as backbone routers. These are in the literature also referred to as ``border routers'' and as such, typically not at the network edge.
\par \textbf{Step 2}: All edge routers with degree above the average degree of all routers in the AS are marked as backbone routers. High-degree routers are unlikely to be access points.
\par The above steps 1 and 2 create a subset $B$ of routers in the graph that represents the set of backbone routers. However, $B$ is not guaranteed to be connected, i.e., subsets in $B$ might be partitioned. In the third step, such partitions are connected to each other, so that a connected set of routers $B$ is established as the network backbone.
\par \textbf{Step 3}: This step runs an algorithm that extends $B$ in order to guarantee a connected backbone for each AS, as listed in Algorithm \ref{alg:connectBackbone}. 
The basic idea of the algorithm is to connect partitions in $B$ by using a breadth-first search (BFS) algorithm. The algorithm starts at an arbitrary router of $B$. Each visited router in the BFS traversal keeps a \emph{parent} field that points to its parent in the BFS tree. If a router visited by the BFS is a backbone router, all the non-backbone predecessors of this router are added to $B$. This way, two partitions of $B$ are connected. In doing so, the algorithm tries to minimize the number of inter-partition routers that are added to $B$ (i.e., find the shortest path between two partitions).

In detail, the algorithm works as follows (cf. Algorithm \ref{alg:connectBackbone}). The initial router is placed in a queue $Q$ holding all routers still to be processed (line 3). For each router $v$ in $Q$, all routers between $v$ and the first router in $B$ according to the predecessor relation are added to $B$ (lines 6 -- 12). This is referred to as the \emph{connection stage}. After the connection stage, the algorithm starts an \emph{expansion stage}, where new neighbors of the connected routers are explored.

In the expansion stage, for each router $v$ in $Q$, the BFS algorithm iterates through $v$'s direct neighborhood (lines 13 -- 20). If a neighbor router $n$ of $v$ is not in $Q$, it is added to $Q$ (lines 14 -- 16). Else, the algorithm tries to optimize the route between $n$ and the backbone, i.e., to minimize the number of hops in the parent relation of $n$. To this end, if $v$ is in $B$, but the parent of $n$ is not in $B$, $v$ becomes the new parent of $n$ (lines 18 -- 20).

After the expansion stage is finished, the next iteration is started by executing the connection stage for all routers in $Q$, etc., until all routers have been visited.

\subsubsection{Placement of Fog Nodes}\label{sec:fogPlacement}
The Fog node placement algorithm determines the Fog node type and location in the network topology such that the application that is deployed on the Fog nodes can serve all edge routers. In particular, the network latency between any edge router and the closest Fog node needs to be within a given latency bound. Further, the algorithm takes into account the Fog configuration file provided by the user. In the Fog configuration file, the specification of a Fog node type indicates the maximum number of clients (i.e., instances of the client application) a Fog node can serve (i.e., its capacity) and its deployment cost (e.g., monetary cost). Further, the configuration file specifies the \emph{edge occupancy} as the average number of clients connected to one edge router (e.g., access point) of the network topology.

The idea of the Fog node placement algorithm, listed in Algorithm \ref{alg:fogPlacement}, is based on a greedy algorithm to find optimal placements for web server replicas in the topology of the Internet, proposed by Qiu et al. \cite{916655}.
The problem of placing Fog nodes is similar to theirs, as both try to serve a number of clients with minimal cost within a latency bound between client and closest replica or Fog node, respectively.

\textproc{PlaceFogNodes} is the function to determine the Fog node placement (lines 1 -- 11). As input parameters, the function requires the set of edge routers $A$ (i.e., all routers not in the set of backbone routers $B$) and a latency threshold $T$ (maximum network latency between any edge router and the closest Fog node).
The algorithm determines a subgraph of \emph{candidate routers} from the original network topology using the function \textproc{DeterminePossibleFogNodes} (lines 12 -- 25). A candidate router is any router within the latency threshold $T$ from any edge router in the topology. Further, the function \textproc{DeterminePossibleFogNodes} assigns deployment costs based on the number and type of Fog nodes needed at the respective candidate router in order to serve the clients of all edge routers in its range; this is computed based on the specifications (Fog node types and edge occupancy) provided in the Fog configuration file (Figure \ref{fig:workflow}). The candidate routers are sorted based on the ratio of deployment costs to \emph{coverage}---i.e., number of edge routers covered in the latency range $T$ from the candidate router. The router $f$ with the highest ratio $\frac{\mathit{coverage}}{\mathit{cost}}$ is added to the set of Fog nodes $F$.
All edge routers that are covered by $f$, i.e., that are in the latency range $T$ from $f$, are removed from $A$. Then, the function \textproc{DeterminePossibleFogNodes} is called on the reduced set of non-covered edge routers $A$. The algorithm terminates when all edge routers are covered.

\begin{algorithm}
\small
	\caption{Fog Node Placement Algorithm}
	\label{alg:fogPlacement}
	\begin{algorithmic}[1]
		\Procedure{PlaceFogNodes}{$G,A,T$}
			\State $F \gets \{\}$
			
			\While{$A \neq \{\}$}
				\State $C \gets \Call{DeterminePossibleFogNodes}{G,A,T}$
                \State $\Call{Sort}{C}$
				\State $f \gets C.\Call{First}{ }$
				\State $F \gets F \cup \{f\}$
				\State $A \gets A\setminus f.\mathit{range}$
			\EndWhile
			\State \textbf{return} $F$
		\EndProcedure
		\Statex

		\Procedure{DeterminePossibleFogNodes}{$G,A,T$}
			\State $C \gets \{\}$
			\ForAll{$a \in A$}
				\ForAll{$v \in G \land \mathit{latency(v,a)} \leq T$}
                	\State $C \gets C \cup \{v\}$
                \EndFor
			\EndFor
            \ForAll{$c \in C$}
            	\State $R(c) \gets$ all routers in range $T$ from $c$
                \State find the cost-optimal Fog node configuration that can serve all routers in $R(c)$
                \State save the range and the deployment cost of the optimal configuration in $c$
            \EndFor
			\State \textbf{return} $C$
		\EndProcedure
	\end{algorithmic}
\end{algorithm}

%% file: content/evaluation.tex
\section{Evaluation}
\label{sec:evaluation}

\begin{figure*}
\begin{minipage}[t]{0.22\linewidth}
	\begin{overpic}[width=\linewidth]{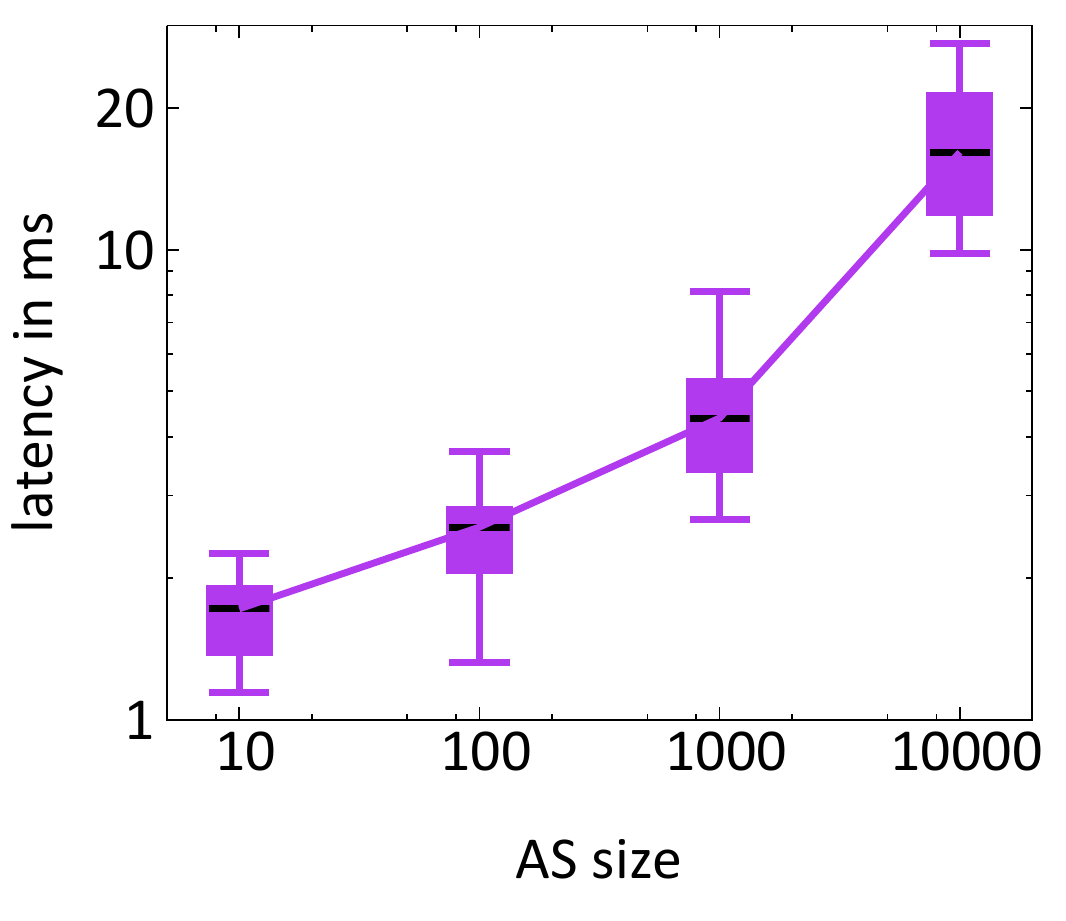}
        		\put(0,0){(a)}
      	\end{overpic}
\end{minipage}%
    \hfill%
\begin{minipage}[t]{0.22\linewidth}
	\begin{overpic}[width=\linewidth]{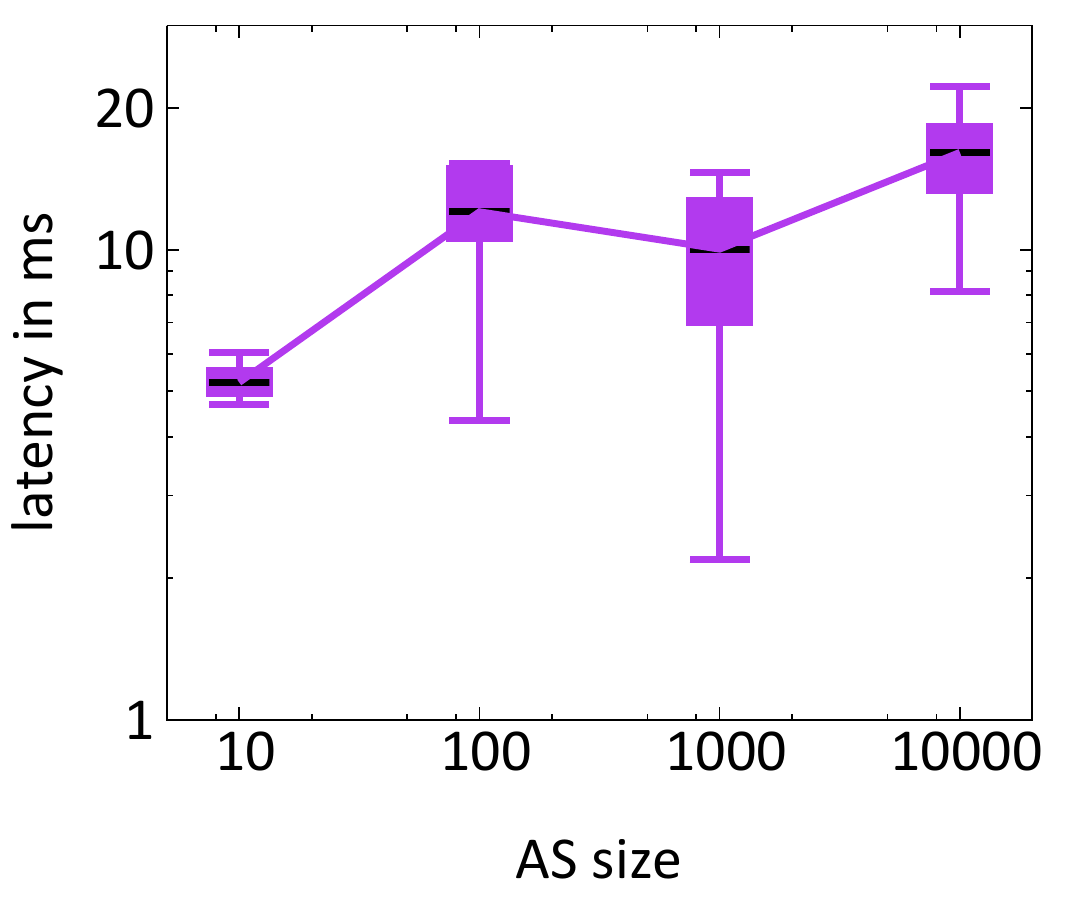}
        		\put(0,0){(b)}
      	\end{overpic}
\end{minipage}%
    \hfill%
\begin{minipage}[t]{0.22\linewidth}
	\begin{overpic}[width=\linewidth]{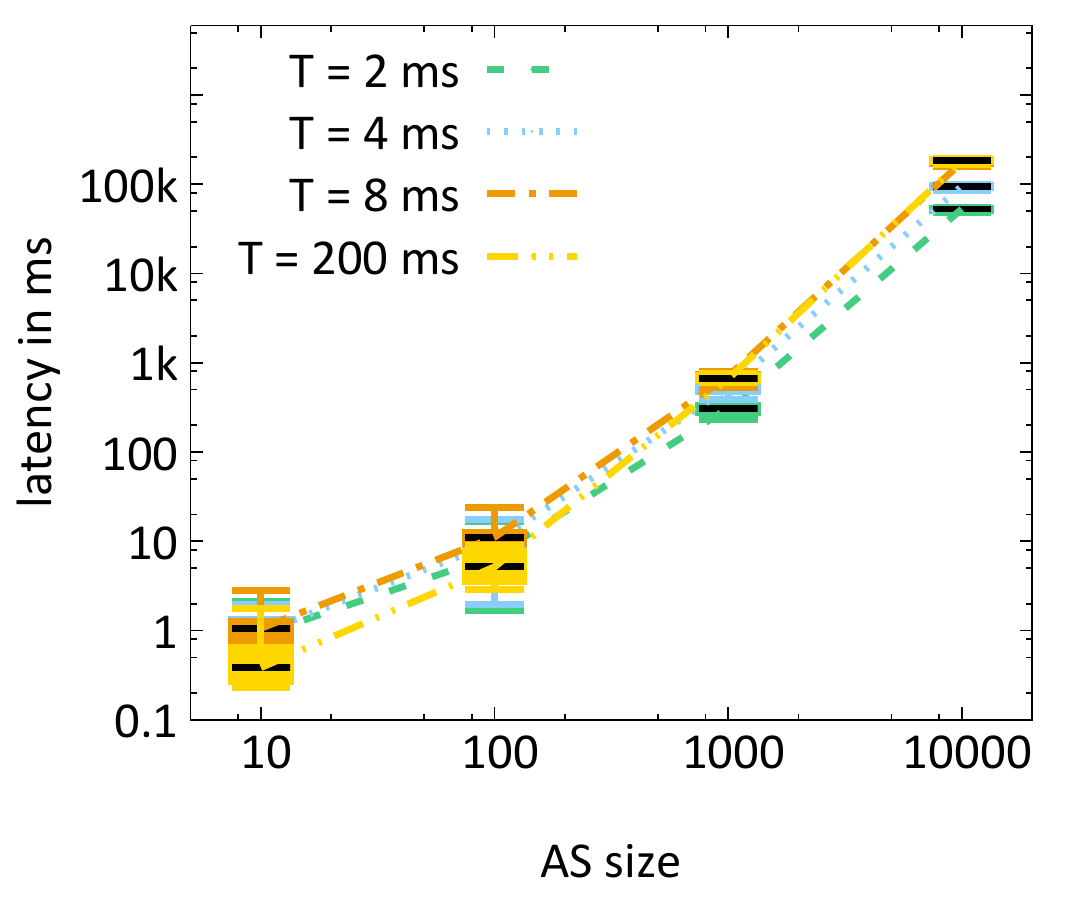}
        		\put(0,0){(c)}
      	\end{overpic}
\end{minipage}%
\hfill
\begin{minipage}[t]{0.22\linewidth}
	\begin{overpic}[width=\linewidth]{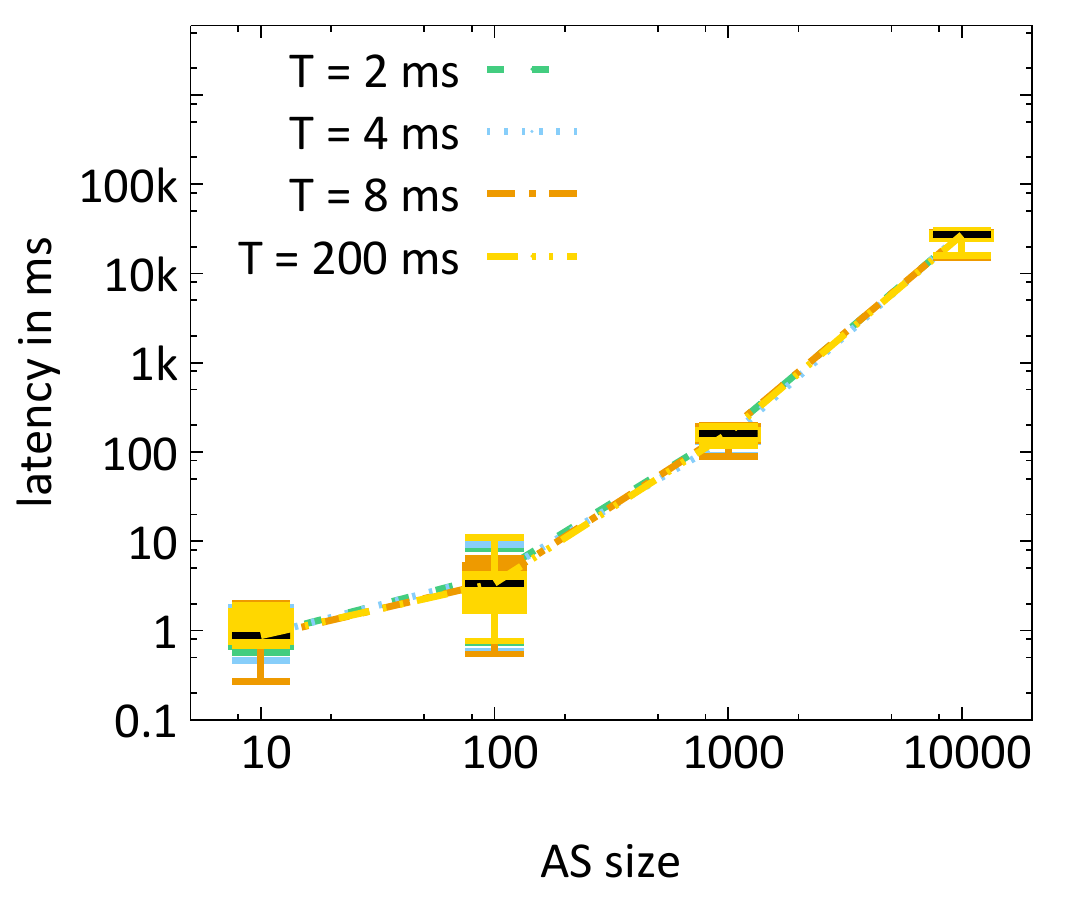}
        		\put(0,0){(d)}
      	\end{overpic}
\end{minipage}%
\hfill
\vspace{-0.2cm}
\caption{Evaluations. (a) Edge identification (BRITE) (b) Edge identification (CAIDA) (c) Fog node placement (BRITE) (d) Fog node placement (CAIDA). Results depict 10th, 25th, 50th, 75th and 90th percentile in ``candlesticks'' representation.}
\label{fig:eval}
\vspace{-0.4cm}
\end{figure*}

\begin{figure}
\begin{minipage}[t]{0.45\linewidth}
	\begin{overpic}[width=\linewidth]{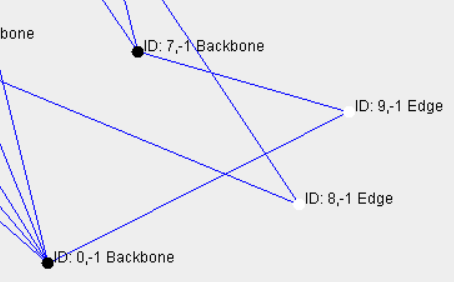}
        		\put(0,-5){(a)}
      	\end{overpic}
\end{minipage}%
\hfill%
\begin{minipage}[t]{0.45\linewidth}
	\begin{overpic}[width=\linewidth]{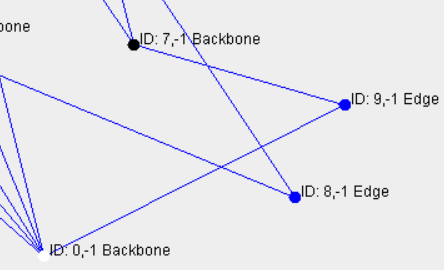}
        		\put(0,-5){(b)}
      	\end{overpic}
\end{minipage}%
    \hfill%
\vspace{-0.0cm}
\caption{Fog node placement. (a) Close to edge. (b) Further from edge. The white circles represent Fog nodes.}
\label{fig:evalFog}
\vspace{-0.3cm}
\end{figure}

Here we present experiments that evaluate the performance and the efficacy of the edge identification and the Fog node placement in \sys.
To show the versatile nature of \sys, we use two different Internet topology datasets:
first, a synthetic topology generated by the BRITE topology generator \cite{Medina:2001:BAU:882459.882563}, using the model of Albert and Barab\'{a}si \cite{PhysRevLett.85.5234};
second, a real-world topology from CAIDA \cite{caida} measured in 2014\footnote{\url{http://data.caida.org/datasets/topology/ark/ipv4/itdk/2014-12/}}.

\subsection{Performance}
For the performance evaluation, we used autonomous systems of different sizes ($n$ = 10, 100, 1000 and 10000 nodes).
Each size is evaluated with five different samples and five runs each.
For the BRITE dataset, the autonomous systems are generated with exactly $n$ nodes. From the CAIDA dataset, we select autonomous systems with a deviation of $\pm 5\%$ from $n$ so that enough different autonomous systems of similar size can be found.
The evaluation was carried out on an Intel i5-4670K processor with 4 phyiscal and logical threads @3.4GHz using 16 GB RAM; the operating system was Ubuntu 17.04.

We implemented adapters for both datasets to generate a generalized topology for edge identification and Fog node placement. In the evaluations, we measure the latency for performing the two major preprocessing steps in \sys: edge identification and Fog node placement.

\subsubsection{Edge Identification}

 As the theoretical complexity of the edge identification is $\mathcal{O}(|V|+|E|)$ (i.e., the time complexity of BFS), we expect a proportional growth of latency to the number of nodes. However, as there can only be a fixed maximum number of edges connected to each node (because a router only has a limited number of ports), the complexity should grow linearly with the number of routers.
 
 The latency for the edge identification are depicted in Fig. \ref{fig:eval} (a) and (b).
For the BRITE topologies (Fig. \ref{fig:eval} (a)), one can clearly see a linearly increasing latency depending on the size of the AS.
For the CAIDA dataset (Fig. \ref{fig:eval} (b)) one can also identify a trend in the increasing time; however, it is not monotonic as with the BRITE topologies. Also the deviation of results is larger. This may be related to issues with the CAIDA dataset, as measuring the Internet topology is a hard and error-prone task.


\subsubsection{Fog node placement}

Since the Fog node placement algorithm depends on the edge-to-fog latency threshold $T$, we evaluated this algorithm using different threshold values.
We evaluated each AS with 2, 4, 8 and 200 ms.
The results are depicted in Fig. \ref{fig:eval} (c) and (d).
For both datasets the running time increases proportional to the number of routers.

The BRITE topologies in Fig. \ref{fig:eval} (c) also show an increasing latency with an increasing threshold $T$ setting.
Even though the latency increases, it does not increase linearly with increasing $T$.
If $T$ exceeds the diameter of the graph, a higher $T$ will not increase the latency of the Fog node placement algorithm, as the number of candidate nodes does not grow further.
This can be seen on smaller graphs and with the higher values $T=8$ and $T=200$ ms. 

Similar to the edge identification, the CAIDA topology results look a bit different.
Despite changing thresholds $T$ the latency barely varies throughout the experiments.
This may be related to the properties of the CAIDA dataset: Routers in the CAIDA topologies have a lower degree than in the topologies generated by BRITE. 

\subsection{Visualization of Topology Enhancement}
Figure \ref{fig:evalFog} shows a small excerpt of a visualization of the Fog topology enhancement algorithm. The intent is to show that the algorithm makes plausible choices in Fog node placement commensurate with the intent of the topology generation criterion (i.e., edge node versus core nodes in the Fog infrastructure).
As can be seen in the figure, the edge identification algorithm has identified low-degree routers as edge routers, while high-degree routers are marked as backbone routers. Depending on the latency bound given, the Fog node placement algorithm either placed two Fog nodes at the edge (Figure \ref{fig:evalFog} (a)), or one Fog node in the backbone (Figure \ref{fig:evalFog} (b)). Whilst being just a small example, this visualization is intended to show the operation of the proposed algorithms for Fog topology generation.


%% file: content/conclusion.tex
\section{Conclusion and Future Work}
\label{sec:conclusion}
In this paper, we have proposed a scalable and extensible emulation framework for Fog Computing environments, called \sys. We highlighted the fundamental components of \sys, discussed the provided algorithms for Fog topology enhancement and its extensible nature. Evaluations show the scalability of the topology enhancement algorithms with respect to topology size and the efficacy of the approach. 
\par In future work, we plan to augment \sys with further capabilities. Embedding mobility models both for clients and Fog nodes would support the evaluation of mobile users that connect to different access points as well as Fog nodes deployed on mobile nodes such as cars or drones \cite{Mayer:2017:FMS:3055601.3055614}. Furthermore, support for hierarchical Fog infrastructures will be added \cite{openfog, hong2013mobile, foglets}.